\documentclass[
               twocolumn,
               noshowpacs,          
               nopreprintnumbers,     
               aps,                 
               prl,                 
               letter,             
	       superscriptaddress,  
               nofootinbib,         
               tightenlines,        
               floats,floatfix    
               ]{revtex4-1}

\usepackage{amsmath, amsfonts, amsthm, amssymb, graphicx}
\usepackage{color}
\usepackage[utf8]{inputenc}
\usepackage{lipsum}
\usepackage{centernot}
\usepackage[normalem]{ulem}

\usepackage{feynmp}

\def\be{\begin{equation}}
\def\ee{\end{equation}}
\def\ba{\begin{eqnarray}}
\def\ea{\end{eqnarray}}
\def\beastar{\begin{eqnarray*}}
\def\eeastar{\end{eqnarray*}}       

\def\del{\partial}
\def\bdm{\begin{displaymath}}
\def\edm{\end{displaymath}}

\def\bq{\begin{quote}}
\def\eq{\end{quote}}

 at 10truept
\newcommand{\ud}{{\rm d}}

\newcommand{\beq}{\begin{equation}}
\newcommand{\eeq}{\end{equation}}
\newcommand{\bea}{\begin{eqnarray}}
\newcommand{\eea}{\end{eqnarray}}
\newcommand{\beqa}{\begin{eqnarray}}
\newcommand{\eeqa}{\end{eqnarray}}

\def\ltap{\ \raise.3ex\hbox{$<$\kern-.75em\lower1ex\hbox{$\sim$}}\ }
\def\gtap{\ \raise.3ex\hbox{$>$\kern-.75em\lower1ex\hbox{$\sim$}}\ }
\def\gl{\ \raise.5ex\hbox{$>$}\kern-.8em\lower.5ex\hbox{$<$}\ }
\def\roughly#1{\raise.3ex\hbox{$#1$\kern-.75em\lower1ex\hbox{$\sim$}}}

\begin{document}

\title{Deconstructing higher order clockwork gravity}
\author{ Anastasios Avgoustidis} 
\email{anastasios.avgoustidis@nottingham.ac.uk}
\affiliation{School of Physics and Astronomy, 
University of Nottingham, Nottingham NG7 2RD, UK} 
\author{Florian Niedermann } 
\email{niedermann@cp3.sdu.dk}
\affiliation{Nordita, KTH Royal Institute of Technology and Stockholm University, Roslagstullsbacken 23, SE-106 91 Stockholm, Sweden}
\author{Antonio Padilla} 
\email{antonio.padilla@nottingham.ac.uk}
\affiliation{School of Physics and Astronomy, 
University of Nottingham, Nottingham NG7 2RD, UK} 
\author{Paul M. Saffin} 
\email{paul.saffin@nottingham.ac.uk}
\affiliation{School of Physics and Astronomy, 
University of Nottingham, Nottingham NG7 2RD, UK}

\date{\today}

\begin{abstract}
We consider the higher order clockwork theory of gravitational interactions, whereby a number of gravitons are coupled together with TeV strength, but nevertheless generate a Planck scale coupling to matter without the need for a dilaton. It is shown that the framework naturally lends itself to a five-dimensional geometry, and we find the 5D continuum version of such deconstructed 4D gravitational clockwork models. Moreover, the clockwork picture has matter coupled to particular gravitons, which in the 5D framework looks like a braneworld model, with the Randall-Sundrum model being a special case. More generally, the gravitational clockwork leads to a family of scalar-tensor braneworld models, where the scalar is not a dilaton.
  \end{abstract}
\maketitle

\section{Introduction}
The large hierarchy of scales between electroweak physics and gravitational physics remains somewhat of a mystery. There are, however, mechanisms that are able to generate such hierarchies. A well-known continuum model was developed by Randall and Sundrum~\cite{RS1}, whereby co-dimension one branes were embedded in an anti-de Sitter (AdS) spacetime, with the result that the effective four dimensional scale of gravity is a combination of the AdS scale and the five-dimensional scale of gravity. In a separate development, clockwork mechanisms were shown to be able to generate exponential hierarchies between the fundamental energy scale and the effective energy scale of the low energy degrees of freedom \cite{cw1,cw2}. This is achieved by having a number of fields coupled together sequentially, where the coupling between each pair of fields in the sequence is ${\cal O}(1)$. Despite having such ${\cal O}(1)$ couplings, if the sequence of fields is long enough one finds a zero mode with an exponentially small energy scale compared to the fundamental scale. In \cite{clocktheory} it was shown that the mechanism also applied to gravitational physics, at the \textit{linear} level, provided couplings were made exponentially site-dependent. In the deconstructed version of the theory, this site dependence translates into a coupling to a dilaton. The framework of \cite{Hinterbichler:2012cn}, which describes how multiple gravity sectors may be coupled together consistently, was used in \cite{cwgravity} to create a model with a discrete number of gravitons, coupled in a clockwork sense. This fully \textit{non-linear} theory of gravity was also able to generate a hierarchy of scales, this time between the single scale appearing in the clockwork action and the Planck scale associated to the massless graviton mode. In particular, this model does not introduce parameter hierarchies at the non-linear level and as a result, it will not require the addition of a dilaton in the continuum limit.

In this paper we follow the deconstruction philosophy of \cite{ArkaniHamed:2003vb} and consider this discrete set of four-dimensional gravitons to simply be the gravitational field of a five-dimension model at discrete locations along some fifth extra dimension. Having made this identification we may take the continuum limit to discover the continuum version of the  higher order gravitational clockwork. Following the clockwork idea of placing matter at the end of the clockwork chain we are led directly to braneworld models, the simplest of which is the Randall-Sundrum model. More generally we find a scalar-tensor model, but one where the scalar is not a dilaton. As alluded to above, this has to be contrasted with the linear gravitational clockwork proposed in \cite{clocktheory}, which was shown to arise from dimensionally deconstructing the linear dilaton model~\cite{LinDilaton}. There, the presence of the 5D dilaton was crucial for generating site-dependent couplings in the 4D theory required to obtain the standard clockwork mass matrix~\cite{Craig}.

In the following we shall start with a brief recap of the four-dimensional gravitational clockwork, then see how to express it in the language of five-dimensional geometry. Having done that we shall see how this naturally leads, in the first instance, to the Randall-Sundrum model and, in more generality, to scalar-tensor braneworld models.

\section{Deconstructing clockwork gravity}  \label{sec:dec}
As described in \cite{cwgravity}, following \cite{Hinterbichler:2012cn}, one may write a four-dimensional multi-gravity theory with $N$ metric fields $g_{(i)}$, with $i=0,1,...,N-1$, in a neat way with kinetic, $S_K$, and potential, $S_V$, terms in the action  given by
\ba
S_K&=&\frac{M_{(4)}^2}{2}\sum_{i=0}^{N-1}\int\ud^4x\sqrt{-g_{(i)}}R_{(i)},\\
S_V&=&\sum_{i,j,k,l}\int T_{ijkl}\epsilon_{abcd}e^{(i)a}\wedge e^{(j)b}\wedge e^{(k)c}\wedge e^{(l)d},
\ea
where the constants $T_{ijkl}$ determine the coupling of the various multigravitons to each other, and  $e^{(i) a} = e^{(i) a}_{\,\,\,\mu} \ud x^\mu$ with the vielbeins defined through $g_{(i)\mu\nu} = e^{(i) a}_{\,\,\,\mu} e^{(i) b}_{\,\,\,\nu}\eta_{ab} $. Going beyond pairwise coupling by including ``loops", such as coupling $e^{(1)}\to e^{(2)}\to e^{(3)}\to e^{(1)}$, generically leads to ghosts \cite{multimetricghosts}\cite{cycles}\cite{rescue}, and so we shall restrict ourselves, as in  \cite{cwgravity}, to nearest neighbour interactions. Given the motivation of this paper, namely to deconstruct the gravitational clockwork model, this is natural, and the nearest neighbour interactions turn into derivatives of five dimensional quantities. This nearest-neighbour restriction leads to 
\begin{align}
S_V&=\sum_i\int \left( T_{iiii}\epsilon_{abcd}e^{(i)a}\wedge e^{(i)b}\wedge e^{(i)c}\wedge e^{(i)d} \right.\\\nonumber
				&+4T_{iii,i+1}\epsilon_{abcd}e^{(i)a}\wedge e^{(i)b}\wedge e^{(i)c}\wedge e^{(i+1)d} \\\nonumber
				&+6T_{ii,i+1,i+1}\epsilon_{abcd}e^{(i)a}\wedge e^{(i)b}\wedge e^{(i+1)c}\wedge e^{(i+1)d} \\\nonumber
				&\left.+4T_{i,i+1,i+1,i+1}\epsilon_{abcd}e^{(i)a}\wedge e^{(i+1)b}\wedge e^{(i+1)c}\wedge e^{(i+1)d}\right)\\\label{eq:SV_nearest_neighbour}\nonumber
	&=\sum_i\int \ud^4x\sqrt{-g^{(i)}}\left( -24T_{iiii} -24T_{iii,i+1}e_\mu^{(i+1)a}e^{(i)\mu}_{\;\;\;\;\;\;a}\right.\\\nonumber
		&-24T_{ii,i+1,i+1}e_{[\mu}^{(i+1)a}e_{\nu]}^{(i+1)b}e^{(i)\mu}_{\;\;\;\;\;\;a}e^{(i)\nu}_{\;\;\;\;\;\;b}\\
			&\left.-24T_{i,i+1,i+1,i+1}e_{[\nu}^{(i+1)a}e_\rho^{(i+1)b}e_{\sigma]}^{(i+1)c}e^{(i)\nu}_{\;\;\;\;\;\;a}e^{(i)\rho}_{\;\;\;\;\;\;b}e^{(i)\sigma}_{\;\;\;\;\;\;c}\right).
\end{align}
Another nice property of taking only nearest neighbour interactions is that the Deser-van Nieuwenhuizen symmetric vierbein condition is guaranteed by the field equations \cite{cycles},
\ba\label{eq:symm_veirbein_cond}
e^{(i)\mu}_{\;\;\;\;\;\; a}e_{\;\;\;\mu}^{(j)\;b}&=&e^{(i)\mu b}e^{(j)}_{\;\;\;\mu a},
\ea
which will be useful later.

The idea now is to relate the terms in (\ref{eq:SV_nearest_neighbour}) to five dimensional quantities. For this, we need a 5D line element, which we take to be
\ba
\ud s^2&=&g_{\mu\nu}(x,y)\ud x^\mu \ud x^\nu+\ud y^2,
\ea
where the $\mu$, $\nu$ indices run over four dimensions. The extrinsic curvature of constant-$y$ surfaces then has the following non-zero components
\ba
K_{\mu\nu}&=&\frac{1}{2}\del_y g_{\mu\nu}(x,y).
\ea
The way we have structured our potential (\ref{eq:SV_nearest_neighbour}) means that it is more useful to express extrinsic curvature quantities in terms of the vierbein, and if we note the usual relation $g_{\mu\nu}=\eta_{ab}e_\mu^{\;\;a}e_\nu^{\;\;b}$ (suppressing the multi-graviton index) then we find expressions such as the following for the trace of $K_{\mu\nu}$,
\ba
K&=&\frac{1}{2}g^{\mu\nu}K_{\mu\nu}=e^{\mu a}\del_y e_{\mu a}.
\ea
The next step in the deconstruction process is to interpret the indices $i$ in (\ref{eq:SV_nearest_neighbour}) as corresponding to a location in the fifth dimension, and we are led to introduce finite difference expressions for derivatives in the $y$ direction,
\ba
\del_y e_{\mu a}(x,y)\to\frac{1}{\delta y}\left[ e^{(i+1)}_{\mu a}(x)-e^{(i)}_{\mu a}(x) \right],
\ea
so the continuum $y$-label is replaced with a discrete $i$-label.

It is then a straightforward, albeit somewhat tedious, process to rewrite (\ref{eq:SV_nearest_neighbour}), using the symmetric vierbein condition (\ref{eq:symm_veirbein_cond}), in the form 
\begin{align}\label{eq:S_5D}
S&=\int \ud^5x\sqrt{-g}\left[ \frac{M_{(5)}^3}{2}R_{(5)}-2\Lambda_{(5)}(y)+\alpha_1(y)M^4_{(5)}K\right.\\\nonumber
	&\qquad\qquad\qquad\left.+\alpha_2(y)M^3_{(5)}K_{(2)}+\alpha_3(y)M^2_{(5)}K_{(3)} \right],
\end{align}
where
\begin{align}
&K_{(2)}=\delta^{[\mu}_\alpha\delta^{\nu]}_\beta K^\alpha_{\;\;\mu} K^\beta_{\;\;\nu},\\
&K_{(3)}=\delta^{[\mu}_\alpha\delta^\nu_\beta\delta^{\rho]}_\gamma K^\alpha_{\;\;\mu} K^\beta_{\;\;\nu}K^\gamma_{\;\;\rho},\\\label{eq:lagrangian_coeffs_from_T_1}
&2\Lambda_{(5)}(y)=\frac{24}{\delta y}\left( T_{iiii}+4T_{iii,i+1}+6T_{ii,i+1,i+1}\right.\\\nonumber
		&\qquad\qquad\qquad\left.+4T_{i,i+1,i+1,i+1} \right),\\\label{eq:lagrangian_coeffs_from_T_2}
&\alpha_1(y)M^4_{(5)}=-24\left(T_{iii,i+1}+3T_{ii,i+1,i+1}-3T_{i,i+1,i+1,i+1}\right),\\\label{eq:lagrangian_coeffs_from_T_3}
&\alpha_2(y)M^3_{(5)}-M^3_{(5)}=-24\delta y\left(T_{ii,i+1,i+1}+2T_{i,i+1,i+1,i+1}\right),\\\label{eq:lagrangian_coeffs_from_T_4}
&\alpha_3(y)M^2_{(5)}=-24\delta y^2T_{i,i+1,i+1,i+1},\\
&M^3_{(5)}=\frac{M^2_{(4)}}{\delta y}.
\end{align}
Note that the above equations may be inverted to find the clockwork couplings $T_{ijkl}$ in terms of the deconstructed functions as 
\begin{align}\label{eq:T_from_lagrangian_coeffs_1}
&24T_{iiii}=2\Lambda_{(5)}\delta y+28\alpha_3\frac{M^2_{(5)}}{\delta y^2}-6\frac{\alpha_2M^3_{(5)}-M_{(5)}^3}{\delta y}\\\nonumber
&\qquad\qquad+4\alpha_1M^4_{(5)},\\\label{eq:T_from_lagrangian_coeffs_2}
&24T_{iii,i+1}=-9\alpha_3\frac{M^2_{(5)}}{\delta y^2}+3\frac{\alpha_2M^3_{(5)}-M_{(5)}^3}{\delta y}-\alpha_1M^4_{(5)},\\\label{eq:T_from_lagrangian_coeffs_3}
&24T_{ii,i+1,i+1}=2\alpha_3\frac{M^2_{(5)}}{\delta y^2}-\frac{\alpha_2M^3_{(5)}-M_{(5)}^3}{\delta y},\\\label{eq:T_from_lagrangian_coeffs_4}
&24T_{i,i+1,i+1,i+1}=-\alpha_3\frac{M^2_{(5)}}{\delta y^2}.
\end{align}

\section{Randall-Sundrum braneworld}  \label{sec:RS}
A key part of the clockwork picture is that matter is not coupled to all of the gravitons, and then the gears of the clockwork act such that the zero mode graviton is weakly coupled to that matter. In the deconstructed framework we are presenting, this has the natural interpretation as a braneworld model, with branes living at discrete locations in the fifth dimension. The classic example of this is the Randall-Sundrum model \cite{RS1}, which we shall now see is included as a special case  in our model.

In general, the full action is  taken to be
\be \label{braneaction}
S=S_\text{bulk} +S_\text{branes},
\ee
where $S_\text{bulk}$ is given by (\ref{eq:S_5D}) and 
\be \label{brane}
S_\text{branes}=\sum_{i=L, R} \int \ud^4 x\sqrt{-\gamma_i}\left\{-\sigma_i+\mathcal{L}_m(\gamma^i_{\mu\nu}, \Psi_i)\right\}.
\ee
The bulk metric is given by $g_{ab}$ with corresponding Ricci scalar, $R_{(5)}$, while the induced metric on each brane is given by $\gamma_{\mu\nu}^i$. The label $i=L, R$, represents the left and the right brane. Matter on the brane, $\Psi_i,$  couples to the induced  metric via the Lagrangian $\mathcal{L}_m$, with the vacuum contribution - or  brane tensions - given explicitly by $\sigma_i$.  To recover the Randall-Sundrum model \cite{RS1},  one may now take the simple case of the clockwork action (\ref{eq:S_5D}) by setting $\Lambda_{(5)}=\text{constant}$, and $\alpha_1=\alpha_2=\alpha_3=0$. 

\section{Clockwork gravity from galileons}  \label{sec:gal}
The clockwork-derived continuum model \eqref{eq:S_5D} contains more than the Randall-Sundrum case, but we note that the presence of extrinsic curvatures generically breaks Poincar\'e invariance in the extra dimension.  However, we can understand this as a spontaneously broken symmetry obtained from higher dimensional Horndeski theories \cite{horn, dgsz}. In particular, we imagine the  Horndeski scalar as a proxy for the coordinate dependence along the extra dimension.  

In this paper, we will focus on a braneworld set-up in which the bulk action  is described by a subset of Horndeski  dubbed kinetic gravity braiding  (KGB) \cite{kgb}. To this end, our generic action is given by (\ref{braneaction})
with  bulk part 
\begin{multline} \label{bulk}
S_\text{bulk}=\int \ud^5 x \sqrt{-g} \Big\{P(\phi, X)-G_3(\phi, X)\square \phi + \\ 
 \frac{M^3_{(5)}}{2} R_{(5)} \Big\}
\end{multline}
We have a single scalar $\phi$ and its kinetic operator \mbox{$X=-\frac12 g^{ab} \partial_a \phi \partial_b \phi$}.  Note that the gravitational coupling in the bulk, $M^3_{(5)}$,  is assumed to be constant and there is no direct coupling between the scalar $\phi$ and the two branes.   These two conditions suggest that the scalar should not be identified with a dilaton. It is in this important sense that we offer a completely new direction to previous work (see e.g \cite{clocktheory, disassembling, comment}). We shall assume $\mathbb{Z}_2$ symmetry across each of the two branes. 

To connect briefly with the deconstructed action \eqref{eq:S_5D}, consider the form of the action \eqref{braneaction}-\eqref{brane}  for the following ansatz
\be \label{ansatz}
\ud s^2=N^2(y)  \ud y^2+a^2(y) \tilde \gamma_{\mu\nu}(x) \ud x^\mu \ud x^\nu, \qquad \phi=\phi(y)
\ee
with the branes located at $y=y_L, y_R$ and $\tilde \gamma_{\mu\nu}(x)$ some four-dimensional metric directed along slices of constant $y$.  
We take ansatz (\ref{ansatz}) for the metric, and for the scalar we take
\be
\phi=\phi(y) \,,
\ee
finding that
\begin{multline}
S_\text{bulk}=\int \ud^5 x \sqrt{-g}  \Big\{ \frac{M^3_{(5)}}{2}R_{(5)}  +P(\phi, X)
\\-G_3(\phi, X)(\phi_{nn}+\phi_n K)  
\Big\} \,,
\end{multline}
where $X=-\frac12 \phi_n^2$ and the suffix $n$ denotes derivatives normal to constant $y$ slices, $\partial_n=\frac1N \partial_y$. As we saw previously, $
K=K^\mu_\mu$ is  the trace of the extrinsic curvature, $K_{\mu\nu}=\frac12 \mathcal{L}_n g_{\mu\nu}$  of constant $y$ slices. For this particular ansatz we have
$
K=4 \frac{a_n}{a}$.  

We can locally choose a gauge such that $N(y)=1$,  in which case the bulk action takes the form of \eqref{eq:S_5D} with $P\left(\phi,-\frac12 \phi'^2\right)-G_3\left(\phi,-\frac12 \phi'^2\right)\phi'' =-2\Lambda_{(5)}(y)$ and $G_3\left(\phi,-\frac12 \phi'^2\right)\phi'=-\alpha_1(y) M^4_{(5)}$.  The higher order couplings including $K_{(2)}$ and $K_{(3)}$ are not generated from our KGB action \eqref{bulk}. They {\it can} be generated from more general Horndeski interactions, although these will also alter the kinetic structure for the graviton. For this reason we focus our attention on the KGB subclass. 

To proceed, we restore the generic gauge choice for the lapse $N(y)$, and note that
\be
\int \ud^5 x \sqrt{-g} \left\{ \ldots \right\}=\int \ud^4 x \sqrt{-\tilde \gamma} \int_{y_L}^{y_R} \ud y Na^4  \left\{ \ldots \right\} \,,
\ee
and in particular 
\begin{multline} \label{EH}
\int \ud^5 x \sqrt{-g} \frac{M^3_{(5)}}{2} R_{(5)}=\\\int \ud^4 x \sqrt{-\tilde \gamma} \int_{y_L}^{y_R} \ud y Na^4 \frac{M^3_{(5)}}{2}  \left\{  \frac{1}{a^2} \tilde R-12 \left( \frac{a_n}{a}\right)^2  -8 \frac{a_{nn} }{a} \right\} \,,
\end{multline}
where $\tilde R$ is the four dimensional Ricci scalar constructed from $\tilde \gamma_{\mu\nu}$.  For the branes, we assume a vacuum configuration since there is only tension, and so our ansatz gives 
\be
S_\text{branes}=-\sum_{i=L, R} \sigma_i  \int \ud^4 x\sqrt{-\tilde \gamma}a_i^4  \,,
\ee
where $a_i=a(y_i)$.

Without loss of generality, we can choose coordinates so that $y_L=0$ and $y_R=l$. When we normalise the scale factor so that $a_L=a(0)=1$, the left brane metric is just given by $\tilde \gamma_{\mu\nu}(x)$.  If $a(y)$ and $N(y)$ satisfy the background equations of motion, the metric \eqref{ansatz} captures both background and zero mode fluctuations. From \eqref{EH},  this allows us to trivially read off the effective four dimensional Planck mass as seen by an observer on the left hand brane. It is given by
\be
M_{(4)}^2=M^3_{(5)}   \int_{0}^{l} \ud y Na^2 \,.
\ee
As in   \cite{GM1}, we focus on vacua with  flat slicings, $\tilde \gamma_{\mu\nu}=\eta_{\mu\nu}$, and  exponential profiles for the lapse function and the warp factor,
\be \label{exp}
a(y)=\exp\left(\frac{2k y}{3}\right), \qquad N(y)=a(y)\exp\left(-2 nk y\right) \,.
\ee
Five dimensional flat space corresponds to $k=0$,  warped anti-de Sitter to $n=\frac13$, and a conformally flat solution to $n=0$.   The existence of these solutions will depend on a judicious choice of $P$ and $G_3$ and will not require us to introduce a dilaton, in contrast to \cite{GM1}. In any event, the four dimensional Planck mass can now be obtained explicitly, 
\be
M_{(4)}^2= \frac{M_{(5)}^3}{2k(1-n) }\left(e^{2kl(1-n)}-1\right) \,.
\ee
Provided $n<1$ and $k>0$,  we get an exponential enhancement of scales and a solution to the hierarchy problem that generalises the one found in \cite{RS1}. We now demonstrate that these profiles can be obtained from simple KGB theories.
\subsection{Cubic galileon}
We begin with a bulk theory described by the cubic galileon \cite{gal},
\be
P(X)=\mu M^3_{(5)} X, \qquad G_3(X)=\nu M_{(5)} X \,,
\ee
 where $\mu, \nu$ are dimensionless constants and we have taken $\phi$ to be dimensionless. The bulk equations of motion now give
 \begin{subequations}
 \begin{align}
 \frac{M^3_{(5)}}{N^2}\left[\mu  \frac{\phi'^2}{2} -6\left(\frac{a'}{a}\right)^2\right]-\frac{M_{(5)}}{N^4} \left[4\nu  \left(\frac{a'}{a} \right) \phi'^3\right] & = 0 \,,\qquad \\
\del_y \left\{a^4 \left[-\frac{M^3_{(5)}}{N} \mu \phi'+\frac{M_{(5)}}{N^3}  {4\nu} \left(\frac{a'}{a} \right) \phi'^2\right] \right\} & = 0 \,,
 \end{align}
  \end{subequations}
 with    the generalised Israel junction conditions \cite{Israel,vish} giving the following boundary conditions at the branes
 \begin{subequations}
 \bea
&&  \left[9 M^3_{(5)}\, \frac{a'}{N \, a} + \nu M_{(5)} \left(\frac{\phi'}{N}\right)^3  \right]_{y=0} \!\!=-\frac{3\, \sigma_L}{2} \,,\\
 &&   \left[9 M^3_{(5)} \,  \frac{a'}{N\,a}  + \nu M_{(5)}  \left(\frac{\phi'}{N}\right)^3 \right]_{y=l}=\frac{3 \, \sigma_R}{2} \,, \\
 &&  \left[-\frac{M^3_{(5)}}{N} \mu \phi'+\frac{M_{(5)}}{N^3}  {4\nu} \left(\frac{a'}{a} \right) \phi'^2\right]_{y=0, l}=0 \,,
 \eea
 \end{subequations}
 where we assumed orbifold boundary conditions for $a(y)$ and $\phi(y)$.
 
 This system is solved by a linear scalar $\phi=q y$ in an anti-de Sitter geometry, $N=1, ~a=e^{\frac{2 k y}{3}}$ (for $y>0$).
 The AdS curvature scale is given by $l_{AdS}^{-1}= 2k/3=M^2_{(5)} \mu/(4\nu q )  $ and the gradient of the scalar by\footnote{We will see later that we need the positive root.} $q=M_{(5)}[-3\mu/(4 \nu^2)]^\frac14$.  The boundary conditions impose a tuning on the tensions, $ \sigma_R=-\sigma_L=8 M^3_{(5)} k/3$, corresponding to slight deformation of the Randall-Sundrum tuning~\cite{RS1}, $\sigma_R = -\sigma_L = 4 M_{(5)}^3 k$.  Notice that a real profile for $\phi$ requires $\mu<0$.  This is the opposite sign to the one we normally take for a canonical scalar and so it will render the trivial, translationally  invariant  vacuum unstable to ghost-like instabilities. However, the vacuum we are interested in here is non-trivial, analogous to the self-accelerating vacua found in four dimensional galileon cosmologies \cite{gal}.  We call it self-AdS, since the anti-de Sitter geometry is inherited from a non-trivial galileon profile, as opposed to a bulk cosmological constant (see \cite{selfads} for simialr ideas in a different context). For $\mu<0$, we can quickly check the stability of the   self-AdS vacuum  by considering scalar fluctuations of the form  $\phi=q \, y+\varphi(y, t)$. To leading order, the bulk Lagrangian now gives 
 \be
\mathcal{L}_\text{eff} =\frac{1}{4}\, \mu \, M_{(5)}^3 \frac{(\del_t \varphi)^2}{a^2}+\ldots \,,
 \ee
which is indeed compatible with stability, although it is understood that a more complete statement requires the inclusion of metric perturbations.
 
 \subsection{Log galileon}
 In order to observe something with a different structure to the Randall-Sundrum case, we now turn our attention to a bulk theory described by the log galileon
 \be
 P(X)=\mu  \,M^3_{(5)} X, \qquad G_3(X)=\nu M^3_{(5)} \log\left( \frac{X}{M^2_{(5)}}\right) \,.
 \ee
 Once again we have $\mu, \nu$ being dimensionless constants.  The field equations in the bulk are now given by
 \begin{subequations}
 \bea
&& \frac{M_{(5)}^3}{N^2}\left[\mu  \frac{\phi'^2}{2} -{6}\left(\frac{a'}{a}\right)^2+8\nu  \phi' \frac{a'}{a} \right]=0 \,,\qquad \\
&&\del_y \left\{a^4 \left[-\frac{M^3}{N}\left( \mu \phi'+8\nu \frac{a'}{a} \right)\right] \right\}=0 \,,
 \eea
 \end{subequations}
 with  the following boundary conditions at the branes 
 \begin{subequations}
 \bea
&&  \left[\frac{6}{N} \frac{a'}{a} -4 \nu \, \frac{\phi'}{N}\right]_{y=0}=-\frac{\sigma_L}{ M^3_{(5)}} \,, \label{eq:log_bdr_1}\\
 &&  \left[\frac{6}{N} \frac{a'}{a} -4 \nu \, \frac{\phi'}{N}\right]_{y=l}=\frac{\sigma_R}{M^3_{(5)}}\,, \label{eq:log_bdr_2}\\
 &&  \left[-\frac{M^3_{(5)}}{N}\left( \mu \phi'+8\nu \frac{a'}{a} \right) \right]_{y=0, l}=0\,.
 \eea
 \end{subequations}
 Provided $\mu=-{16}\nu^2/3$, we can obtain the whole family of solutions given by \eqref{exp}, alongside a linear profile for the scalar $\phi=qy$. The curvature scale in the metric is related to the scalar gradient as $\frac{2 k}{3}=-\mu q/(8 \nu)$. There is no further restriction on the choice of $q$ and $n$ suggesting a multi parameter family of solutions with enhanced symmetry.   Using \eqref{eq:log_bdr_1} and \eqref{eq:log_bdr_2}, we find that the brane tensions have to vanish, $\sigma_{L} = \sigma_{R} = 0$, which corresponds again to a tuning. 
 
We also note that once again the kinetic term for the scalar has the ``wrong sign" around $\phi=$ constant vacua.  As before, this is of no concern to us since the vacuum of interest has a non-trivial scalar gradient.  To quickly check if our vacuum is stable against scalar ghosts, we again  consider fluctuations of the form  $\phi=qy+\varphi(y, t)$. To leading order, the bulk Lagrangian gives 
 \be
\mathcal{L}_\text{eff} =\frac{1}{4}\, \mu \, M_{(5)}^3 \frac{(\del_t \varphi)^2}{a^2}+\ldots \,,
 \ee  which as before is compatible with stability as $\mu < 0$. 
 
\section{Conclusions} \label{sec:conc}
We have constructed the continuum version of the clockwork gravity model presented in \cite{cwgravity}. It was shown that when deconstructed from four to five dimensions, the theory consists of an action containing a standard Ricci scalar term, along with various powers of the extrinsic curvature (\ref{eq:S_5D}). This means that explicit Poincar\'e symmetry is apparently lost in the general higher dimensional model. However, when viewed as a spontaneous breaking we see that the deconstructed model is a (non-dilaton) scalar-tensor theory, and the vacuum profile of the scalar is what leads to the appearance of the extrinsic curvature terms. 

The philosophy of the clockwork paradigm, whereby matter is placed at only a few sites, translates naturally to the braneworld picture in the continuum five-dimensional model. The Randall-Sundrum model was discovered to be included in the deconstructed gravitational clockwork as the simplest case, but we also examined a Kinetic Gravity Braiding model consisting of a standard kinetic term along with a cubic and log galileon interaction, confirming that there are regions of parameter space where a hierarchy of scales is generated. A first check of the stability of the scalar sector has been performed for both the cubic and log galileon case. A full discussion of the stability of the radion field, controlling the size of the extra dimension, is left for future work and would likely involve a stabilisation mechanism \`a la Goldberger-Wise~\cite{GW}.  It would also be interesting to study the phenomenology of these models and the implications for the search for new physics at colliders.

\begin{acknowledgments}
\vskip.5cm

{\bf Acknowledgments}: 
  A.A, A.P. and P.M.S. acknowledge support from the UK Science and Technology Facilities Council through grant ST/S000437/1. A.P. is also funded by a Leverhulme Trust Research Project Grant. FN is supported by Villum Fonden grant 13384. 
\end{acknowledgments}

\end{document}